\input amssym.tex


\magnification=\magstephalf
\hsize=14.0 true cm
\vsize=19 true cm
\hoffset=1.0 true cm
\voffset=2.0 true cm

\abovedisplayskip=12pt plus 3pt minus 3pt
\belowdisplayskip=12pt plus 3pt minus 3pt
\parindent=1.0em


\font\sixrm=cmr6
\font\eightrm=cmr8
\font\ninerm=cmr9

\font\sixi=cmmi6
\font\eighti=cmmi8
\font\ninei=cmmi9

\font\sixsy=cmsy6
\font\eightsy=cmsy8
\font\ninesy=cmsy9

\font\sixbf=cmbx6
\font\eightbf=cmbx8
\font\ninebf=cmbx9

\font\eightit=cmti8
\font\nineit=cmti9

\font\eightsl=cmsl8
\font\ninesl=cmsl9

\font\sixss=cmss8 at 8 true pt
\font\sevenss=cmss9 at 9 true pt
\font\eightss=cmss8
\font\niness=cmss9
\font\tenss=cmss10

 at 12 true pt
 at 12 true pt
\font\bigrm=cmr10 at 12 true pt
 at 12 true pt
 at 12 true pt

 at 16 true pt
 at 16 true pt
\font\Bigrm=cmr12 at 16 true pt
 at 16 true pt
 at 16 true pt

\catcode`@=11
\newfam\ssfam

\def\tenpoint{\def\rm{\fam0\tenrm}%
    \textfont0=\tenrm \scriptfont0=\sevenrm \scriptscriptfont0=\fiverm
    \textfont1=\teni  \scriptfont1=\seveni  \scriptscriptfont1=\fivei
    \textfont2=\tensy \scriptfont2=\sevensy \scriptscriptfont2=\fivesy
    \textfont3=\tenex \scriptfont3=\tenex   \scriptscriptfont3=\tenex
    \textfont\itfam=\tenit                  \def\it{\fam\itfam\tenit}%
    \textfont\slfam=\tensl                  \def\sl{\fam\slfam\tensl}%
    \textfont\bffam=\tenbf \scriptfont\bffam=\sevenbf
    \scriptscriptfont\bffam=\fivebf
                                            \def\bf{\fam\bffam\tenbf}%
    \textfont\ssfam=\tenss \scriptfont\ssfam=\sevenss
    \scriptscriptfont\ssfam=\sevenss
                                            \def\ss{\fam\ssfam\tenss}%
    \normalbaselineskip=13pt
    \setbox\strutbox=\hbox{\vrule height8.5pt depth3.5pt width0pt}%
    \let\big=\tenbig
    \normalbaselines\rm}

\def\ninepoint{\def\rm{\fam0\ninerm}%
    \textfont0=\ninerm      \scriptfont0=\sixrm
                            \scriptscriptfont0=\fiverm
    \textfont1=\ninei       \scriptfont1=\sixi
                            \scriptscriptfont1=\fivei
    \textfont2=\ninesy      \scriptfont2=\sixsy
                            \scriptscriptfont2=\fivesy
    \textfont3=\tenex       \scriptfont3=\tenex
                            \scriptscriptfont3=\tenex
    \textfont\itfam=\nineit \def\it{\fam\itfam\nineit}%
    \textfont\slfam=\ninesl \def\sl{\fam\slfam\ninesl}%
    \textfont\bffam=\ninebf \scriptfont\bffam=\sixbf
                            \scriptscriptfont\bffam=\fivebf
                            \def\bf{\fam\bffam\ninebf}%
    \textfont\ssfam=\niness \scriptfont\ssfam=\sixss
                            \scriptscriptfont\ssfam=\sixss
                            \def\ss{\fam\ssfam\niness}%
    \normalbaselineskip=12pt
    \setbox\strutbox=\hbox{\vrule height8.0pt depth3.0pt width0pt}%
    \let\big=\ninebig
    \normalbaselines\rm}

\def\eightpoint{\def\rm{\fam0\eightrm}%
    \textfont0=\eightrm      \scriptfont0=\sixrm
                             \scriptscriptfont0=\fiverm
    \textfont1=\eighti       \scriptfont1=\sixi
                             \scriptscriptfont1=\fivei
    \textfont2=\eightsy      \scriptfont2=\sixsy
                             \scriptscriptfont2=\fivesy
    \textfont3=\tenex        \scriptfont3=\tenex
                             \scriptscriptfont3=\tenex
    \textfont\itfam=\eightit \def\it{\fam\itfam\eightit}%
    \textfont\slfam=\eightsl \def\sl{\fam\slfam\eightsl}%
    \textfont\bffam=\eightbf \scriptfont\bffam=\sixbf
                             \scriptscriptfont\bffam=\fivebf
                             \def\bf{\fam\bffam\eightbf}%
    \textfont\ssfam=\eightss \scriptfont\ssfam=\sixss
                             \scriptscriptfont\ssfam=\sixss
                             \def\ss{\fam\ssfam\eightss}%
    \normalbaselineskip=10pt
    \setbox\strutbox=\hbox{\vrule height7.0pt depth2.0pt width0pt}%
    \let\big=\eightbig
    \normalbaselines\rm}

\def\tenbig#1{{\hbox{$\left#1\vbox to8.5pt{}\right.\n@space$}}}
\def\ninebig#1{{\hbox{$\textfont0=\tenrm\textfont2=\tensy
                       \left#1\vbox to7.25pt{}\right.\n@space$}}}
\def\eightbig#1{{\hbox{$\textfont0=\ninerm\textfont2=\ninesy
                       \left#1\vbox to6.5pt{}\right.\n@space$}}}

\font\sectionfont=cmbx10
\font\subsectionfont=cmti10

\def\figurecaptionfont{\ninepoint}
\def\tablecaptionfont{\ninepoint}


\newcount\equationno
\newcount\bibitemno
\newcount\figureno
\newcount\tableno

\equationno=0
\bibitemno=0
\figureno=0
\tableno=0


\footline={\ifnum\pageno=0{\hfil}\else
{\hss\rm\the\pageno\hss}\fi}


\def\section #1 \par
{\vskip3ex
\global\def\equationlabel{#1}
\immediate\write\terminal{Section #1.}}


\def\subsection #1 \par
{\vskip0pt plus 0.8 true cm\penalty-50 \vskip0pt plus-0.8 true cm
\vskip2.5ex plus 0.1ex minus 0.1ex
\leftline{\subsectionfont #1}\par
\immediate\write\terminal{Subsection #1}
\vskip1.0ex plus 0.1ex minus 0.1ex
\noindent}


\def\appendix #1. #2 \par
{\vskip0pt plus .20\vsize\penalty-100 \vskip0pt plus-.20\vsize
\vskip 1.6 true cm plus 0.2 true cm minus 0.2 true cm
\global\def\equationlabel{\hbox{\rm#1}}
\global\equationno=0
\leftline{\sectionfont Appendix #1. #2}\par
\immediate\write\terminal{Appendix #1. #2}
\vskip 0.7 true cm plus 0.1 true cm minus 0.1 true cm
\noindent}



\def\equation#1{$$\displaylines{\qquad #1}$$}
\def\enum{\global\advance\equationno by 1
\hfill\llap{{\rm(\the\equationno)}}}

\def\noenum{\hfill}
\def\next#1{\cr\noalign{\vskip#1}\qquad}


\def\ifundefined#1{\expandafter\ifx\csname#1\endcsname\relax}

\def\ref#1{\ifundefined{#1}?\immediate\write\terminal{unknown reference
on page \the\pageno}\else\csname#1\endcsname\fi}

\newwrite\terminal
\newwrite\bibitemlist

\def\bibitem#1#2\par{\global\advance\bibitemno by 1
\immediate\write\bibitemlist{\string\def
\expandafter\string\csname#1\endcsname
{\the\bibitemno}}
\item{[\the\bibitemno]}#2\par}

\def\beginbibliography{
\vskip0pt plus .15\vsize\penalty-100 \vskip0pt plus-.15\vsize
\vskip 1.2 true cm plus 0.2 true cm minus 0.2 true cm
\centerline{\sectionfont References}\par
\immediate\write\terminal{References}
\immediate\openout\bibitemlist=biblist
\frenchspacing\parindent=1.8em
\vskip 0.5 true cm plus 0.1 true cm minus 0.1 true cm}

\def\endbibliography{
\immediate\closeout\bibitemlist
\nonfrenchspacing\parindent=1.0em}


\def\figurecaption#1{\global\advance\figureno by 1
\narrower\figurecaptionfont
Fig.~\the\figureno. #1}

\def\tablecaption#1{\global\advance\tableno by 1
\vbox to 0.5 true cm { }
\centerline{\tablecaptionfont%
Table~\the\tableno. #1}
\vskip-0.4 true cm}

\tenpoint

\immediate\openin\bibitemlist=biblist
\ifeof\bibitemlist\immediate\closein\bibitemlist
\else\immediate\closein\bibitemlist
\input biblist \fi


\def\thismonth{\ifcase\month\or
January\or February\or March\or April\or May\or June\or
July\or August\or September\or October\or November\or December\fi}




\def\rmd{{\rm d}}

\def\rmO{{\rm O}}



\def\proof{\noindent{\sl Proof:}\kern0.6em}

\def\frac#1#2{\hbox{$#1\over#2$}}
\def\dual{\mathstrut^*\kern-0.1em}

\def\lvec#1{\setbox0=\hbox{$#1$}
    \setbox1=\hbox{$\scriptstyle\leftarrow$}
    #1\kern-\wd0\smash{
    \raise\ht0\hbox{$\raise1pt\hbox{$\scriptstyle\leftarrow$}$}}
    \kern-\wd1\kern\wd0}
\def\rvec#1{\setbox0=\hbox{$#1$}
    \setbox1=\hbox{$\scriptstyle\rightarrow$}
    #1\kern-\wd0\smash{
    \raise\ht0\hbox{$\raise1pt\hbox{$\scriptstyle\rightarrow$}$}}
    \kern-\wd1\kern\wd0}
\def\slash#1{\setbox2=\hbox{$\displaystyle#1$}%
             \setbox3=\hbox{$\displaystyle/$}%
             #1\kern-0.8\wd2/\kern-1.0\wd3\kern0.8\wd2\kern0.5pt}


\def\nabstar#1{{\nabla\kern0.5pt\smash{\raise 4.5pt\hbox{$\ast$}}
               \kern-5.5pt_{#1}}}
\def\drv#1{{\partial_{#1}}}
\def\drvstar#1{{\partial\kern0.5pt\smash{\raise 4.5pt\hbox{$\ast$}}
               \kern-6.0pt_{#1}}}

\def\ldrvstar#1{{\lvec{\,\partial}\kern-0.5pt\smash{\raise 4.5pt\hbox{$\ast$}}
               \kern-5.0pt_{#1}}}




\def\psibar{\overline{\psi}}


\def\dirac#1{\gamma_{#1}}
\def\diracstar#1#2{
    \setbox0=\hbox{$\gamma$}\setbox1=\hbox{$\gamma_{#1}$}
    \gamma_{#1}\kern-\wd1\kern\wd0
    \smash{\raise4.5pt\hbox{$\scriptstyle#2$}}}


\def\SUn{{\rm SU}(N)}
\def\tr{{\rm tr}}

\def\Ad{{\rm Ad}\kern0.1em}


\def\varza{{\cal Z}_A}
\def\varzqa{{\cal Z}_{qA}}


\def\Nf{N_{\rm f}}
\def\Nb{N_{\rm b}}
\def\chifl{\chi_t^{\rm fl}}
\def\chidc{\chi_t^{\rm dc}}
\def\SF{S_{\rm F}}

\def\drvS#1{\partial\kern0.5pt\smash{\raise 4.5pt\hbox{$\scriptstyle S$}}
               \kern-6.0pt_{#1}}
\def\drvP#1{\partial\kern0.5pt\smash{\raise 4.5pt\hbox{$\scriptstyle P$}}
               \kern-6.0pt_{#1}}

\def\Mbar{M_m}
\def\Mt{\tilde{M}}

\def\F{F}
\def\FR{\hat{F}}
\def\phiR#1{\phi_{#1}}
\def\qRGI{q_{\hbox{\sixrm RGI}}}

\vbox{\vskip0.0cm}
\rightline{CERN-TH-2021-132}

\vskip 1.3cm
\centerline{\Bigrm On the chiral anomaly and the Yang--Mills gradient flow}
\vskip 0.6 true cm
\centerline{\bigrm Martin L\"uscher$^{\kern1pt\rm a,b}$}

\vskip1.5ex
\centerline{{\it $^{\rm a}$CERN,
Theoretical Physics Department, 1211 Geneva 23, Switzerland}}

\vskip1.0ex
\centerline{{\it $^{\rm b}$Albert Einstein Center for Fundamental Physics}}
\centerline{{\it Institute for Theoretical Physics,
Sidlerstrasse 5, 3012 Bern, Switzerland}}

\vskip 0.9 true cm
\ninepoint
\centerline{\bf Abstract}
\vskip 2.0ex
There are currently two singularity-free universal expressions for
the topological susceptibility in QCD, one based on the Yang--Mills
gradient flow and the other on density-chain correlation
functions. While the latter link the susceptibility to the anomalous
chiral Ward identities, the gradient flow permits the emergence
of the topological sectors in lattice QCD to be understood.
Here the two expressions are shown to coincide in
the continuum theory, for any number of quark flavours in
the range where the theory is asymptotically free.

\tenpoint
\vskip0.7cm

\section 1

{\bf 1.}~The usefulness of the Yang--Mills gradient
flow in QCD largely rests on the
still somewhat surprising
fact that it does not require
renormalization [\ref{WilsonFlow}-\ref{SuzukiRen}].
In particular, correlation functions of
local gauge-invariant fields built from the gauge field,
such as the Yang--Mills action density $s(t,x)$ and
the topological charge density $q(t,x)$,
are finite and have no short-distance singularities
if evaluated at positive gradient-flow time $t$.

An early application of the flow showed that
the field space in a finite space-time volume $V$
dynamically divides into disconnected sectors of
fixed topological charge
\equation{
  Q_t=\int\rmd^4 x\,q(t,x)
  \enum
}
in the continuum limit of lattice QCD [\ref{WilsonFlow}].
Moreover, as long as $t>0$, the sectors are
independent of the flow time and so is the topological susceptibility
$\chifl=\langle Q_t^2\rangle/V$.
The expression provides a regularization-independent
definition of the susceptibility, which is close to
its naive definition, but properly
deals with the fluctuations of the gauge field
and the associated ultraviolet singularities.

Other singularity-free expressions for the susceptibility,
the so-called {\it density-chain formulae}, have
previously been derived using a combination of
chiral Ward identities [\ref{ChiWII}] or the
index theorem [\ref{ChiDC}]. A diagonal
quark mass matrix $M$ is assumed in this case,
with positive entries $m_1,m_2,\ldots$ on
the diagonal, and the susceptibility
is represented through correlation functions like [\ref{ChiDC}]
\equation{
  \chidc=m_1\ldots m_5\int\rmd^4x_1\ldots\rmd^4x_4\,
  \langle P_{12}(x_1)S_{23}(x_2)S_{31}(x_3)P_{45}(x_4)S_{54}(0)\rangle
  \enum
}
of the
scalar and pseudo-scalar quark densities
$S_{rs}(x)$ and $P_{rs}(x)$
with flavour indices $r,s$ chosen in a particular way.
Equation (2) involves five flavours of quarks, which can be
valence quarks, if the theory includes a smaller number
of sea quarks.

\section 2

{\bf 2.}~The gradient-flow and the density-chain expression
for the topological susceptibility, $\chifl$ and $\chidc$, are both well
defined and unambiguously normalized, but they
refer to different aspects of the theory,
the first characterizing the distribution of the topological
charge $Q_t$ in the functional integral, while the second
derives from an essentially algebraic property of the theory
(the anomalous chiral symmetry).
So far their equality has been proved
in the pure $\SUn$  gauge theory [\ref{CeEtAl}]
and numerical simulations suggest
that the expressions coincide in QCD with non-zero numbers of
sea quarks too [\ref{ETMTop}].

In this letter, the equality of the two expressions
is first established assuming
the number $\Nf$ of sea quarks is at least $5$ and such that
asymptotic freedom is not lost.
The theory with fewer flavours of sea quarks
is then briefly discussed in sect.~7.
As an intermediate regularization of the theory, a formulation
of lattice QCD is chosen
(the same as in ref.~[\ref{ChiDC}]),
which preserves chiral symmetry [\ref{GW}-\ref{GWreview}].
Since $\chifl$ and $\chidc$ are both
regularization-independent, the details of the
intermediate regularization are irrelevant,
but the argumentation becomes particularly transparent with
this choice of regularization.
The existence of the continuum limit at the non-perturbative
level and the validity of the general principles
of renormalization are however taken for granted.

Although the quark mass matrix $M$ is eventually set to
a positive diagonal matrix, it is initially
taken to be a general complex matrix.
In the formal continuum theory, the quark action is then
\equation{
  \SF=\int\rmd^4x\,\psibar(x)
  \left\{\slash{D}+MP_{-}+M^{\dagger}P_{+}\right\}\psi(x),
  \qquad P_{\pm}=\frac{1}{2}(1\pm\dirac{5}),
  \enum
}
where the quark fields $\psibar(x),\psi(x)$ carry
a (suppressed) flavour index $r=1,\ldots,\Nf$.
There is, in this case, no need to include
the usual term, $i\theta Q$, proportional
to the topological charge $Q$ in the QCD action,
since it can be traded for
a phase factor of the mass matrix through an anomalous chiral
transformation.

\section 3

{\bf 3.}~The chiral symmetry of the lattice theory
implies that the mass matrix and the quark densities
renormalize multiplicatively.
Moreover, the renormalization constants of the densities
may be set to the inverse of the one of the mass matrix.
In the following, a massless renormalization scheme is assumed and
$M$, $S_{rs}(x)$ and $P_{rs}(x)$ denote
the renormalized mass matrix and densities.

On the lattice, the QCD partition function $Z(M)$ is a well-defined
function of $M$ and so is the free energy
\equation{
  \F(M)=-\ln\{Z(M)\}.
  \enum
}
Any connected correlation function
of the quark densities at zero momentum can be obtained
from the free energy
through repeated differentiation with respect to the real and imaginary
parts of the mass matrix.
Particularly useful in this
context are the differential operators $\drvS{rs}$ and $\drvP{rs}$
implicitly defined by the equations
\equation{
  \drvS{rs}\SF=a^4\sum_x S_{rs}(x),
  \qquad
  \drvP{rs}\SF=a^4\sum_x P_{rs}(x).
  \enum
}
An example of a correlation function
derived from the free energy is then
\equation{
  \drvP{12}\drvS{23}\drvS{31}\F(M)=
  a^{12}\sum_{x_1,x_2,x_3}
  \langle P_{12}(x_1)S_{23}(x_2)S_{31}(x_3)\rangle_{\rm c}
  \enum
}
where $a$ is the lattice spacing and $\langle\ldots\rangle_{\rm c}$
denotes the connected part of the lattice QCD expectation
value $\langle\ldots\rangle$.

The free energy $\F(M)$ is invariant under parity
$M\to M^{\dagger}$ and charge conjugation $M\to M^T$.
Under infinitesimal vector and axial-vector flavour
transformations of the mass matrix by an arbitrary anti-Hermitian
flavour matrix $\lambda$,
\equation{
  \delta^V_{\lambda}M=[\lambda,M],
  \qquad
  \delta^A_{\lambda}M=\{\lambda,M\},
  \enum
}
it transforms according to
\equation{
  \delta^V_{\lambda}\F(M)=0,
  \qquad
  \delta^A_{\lambda}\F(M)=-2\,\tr\{\lambda\}\langle Q\rangle.
  \enum
}
A second application of the axial transformation,
\equation{
  \delta^A_{\eta}\langle Q\rangle=
  2\,\tr\{\eta\}\langle Q^2\rangle_{\rm c},
  \enum
}
then links the free energy to the topological susceptibility.
Equations (8) and (9) follow, purely algebraically, from
the definition and chiral symmetry of the lattice theory
[\ref{ExactChSym}]. In particular, the
chiral anomaly [the term proportional to $\langle Q\rangle$ in eq.~(8)]
derives from the transformation behaviour
of the fermion integration measure.

The lattice version of the
density-chain correlation function on the right of
eq.~(2) is exactly equal to $\langle Q^2\rangle/V$
for any non-zero value of the lattice spacing [\ref{ChiDC}].
If both sides of this equation are expressed through
the free energy, the relation
\equation{
  {1\over6}\bigl\{\delta^A_{\lambda}\delta^A_{\eta}\F(M)\bigr\}_{M=\Mbar}
  =m_1\ldots m_5
  \bigl\{
  \drvP{12}\drvS{23}\drvS{31}\drvP{45}\drvS{54}\F(M)
  \bigr\}_{M=\Mbar}
  \enum
}
is obtained, where $\Mbar={\rm diag}(m_1,\ldots,m_{\Nf})$ and
\equation{
  \lambda={i\over2}{\rm diag}(1,1,1,0,\ldots,0),
  \qquad
  \eta={i\over2}{\rm diag}(0,0,0,1,1,0,\ldots,0),
  \enum
}
are diagonal matrices, the latter
acting on the first three and next two
quark flavours, respectively.
It may be worth mentioning in passing that eq.~(10)
is a consequence of the chiral symmetry properties
of the free energy, i.e.~of eqs.~(7)-(9),
but the proof of this statement is too long to be included here.

\section 4

{\bf 4.}~Correlation functions of the quark densities
in general develop non-integrable short-distance singularities
in the continuum limit.
The free energy $\F(M)$ therefore requires additive renormalization
by a counterterm $\Delta\F(M)$ before the limit can be taken.
Power counting implies that the counterterm
must be a polynomial of degree at most $4$
in the matrix elements of $M$ and $M^{\dagger}$. Moreover,
since the free energy and hence its divergent parts
are invariant under the non-anomalous flavour
symmetries, the counterterm may be chosen to be invariant too.

Taking these remarks into account, the counterterm must be
of the form
\equation{
  \Delta\F(M)=V\biggl[{k_0\over a^4}+
  {k_1\over a^2}\kern0.5pt\tr\{M^{\dagger}M\}+
  k_2\kern0.5pt\tr\{M^{\dagger}M\}^2+k_3\kern0.5pt\tr\{(M^{\dagger}M)^2\}
  \biggr]
  \enum
}
with some
coefficients $k_0,\ldots,k_3$ that depend on the gauge
coupling and the renormalization scale in units of the lattice spacing.
Since $\Nf\geq5$ is assumed,
a divergent term proportional to the real part of $\det M$ is
excluded and this in turn implies that
the renormalized free energy
\equation{
  \FR(M)=\F(M)+\Delta\F(M)
  \enum
}
transforms like $\F(M)$
under both singlet and non-singlet flavour transformations.
In particular,
\equation{
  \delta^A_{\lambda}\FR(M)=-2\,\tr\{\lambda\}\langle Q\rangle,
  \enum
}
which shows that $\langle Q\rangle$ does not require renormalization.

Since $\Delta\F(M)$ is annihilated by the axial variations
$\delta^A_{\lambda}$ as well as by any product of more than $4$ derivatives
with respect to the elements of the mass matrix,
the unrenormalized free energy may be replaced by the renormalized one
on both sides of eq.~(10).
The continuum limit can then be taken and, recalling eq.~(2),
this leads to the representation
\equation{
  \chidc=
  {1\over6V}
  \bigl\{\delta^A_{\lambda}\delta^A_{\eta}\FR(M)\bigr\}_{M=\Mbar}
  \enum
}
of $\chidc$, where $\lambda$ and $\eta$ are as in eq.~(11).

\section 5

{\bf 5.}~Contact with the topological charge
at positive gradient-flow time is now made
in the continuum theory
through the small flow-time expansion
\equation{
  q(t,x)\mathrel{\mathop\sim_{t\to0}}
  \sum_{k=1}^{\infty}c_k(t)\phiR{k}(x)
  \enum
}
of the flowed topological charge density
[\ref{RenFlow},\ref{CeEtAl},\ref{Mainz},\ref{HiedaSuzuki}].
The fields on the right of eq.~(16)
are renormalized local fields
at flow time $t=0$ of increasing dimension $d_k$.
As $t\to0$ and up to logarithmically varying factors,
the coefficients $c_k(t)$ scale proportionally to
$t^{d_k/2-2}$.

The fields $\phiR{k}(x)$ must be polynomials
in the elements of the mass matrix $M$,
the basic fields and their derivatives.
They must have the same symmetry properties as the charge density,
where it is understood that $M$ is treated as a ``spurion''.
Fields satisfying these conditions have dimension $d_k\geq4$
and the ones of dimension $4$ are
\equation{
  \phiR{1}(x)=q(x)+\varzqa\drv{\mu}A^s_{\mu}(x),
  \enum
  \next{2.0ex}
  \phiR{2}(x)=\varza\drv{\mu}A^s_{\mu}(x),
  \enum
}
$q(x)$ being the bare charge density, $A^s_{\mu}(x)$ the bare
flavour-singlet axial current, $\varzqa$ a (divergent) mixing
coefficient and $\varza$ the axial-current
renormalization constant.
The mass term
\equation{
  \sum_{r,s}
  \bigl\{
  (M^{\dagger}-M)_{rs}S_{rs}(x)+
  (M^{\dagger}+M)_{rs}P_{rs}(x)
  \bigr\}
  \enum
}
and the ``field'' $\det M^{\dagger}-\det M$ have the required
symmetry properties too, but the latter has dimension $\Nf\geq5$
and the mass term is omitted, because it is linearly related to
the fields (17),(18) through the singlet axial-current
conservation equation.
On the lattice, the absence of a multiplicative
renormalization of the charge density in eq.~(17)
is implied by the chiral Ward identities [\ref{GiustiEtAlI}],
while in the case of dimensional regularization it was recently shown to
derive from the special algebraic properties of the
charge density [\ref{DRqtop}].

The exact asymptotic behaviour of the coefficients
$c_1(t)$ and $c_2(t)$ near $t=0$ can be worked
out in perturbation theory [\ref{RenFlow},\ref{Mainz}].
To this end, they
are first expanded in powers of the renormalized gauge coupling $g$
and the series is then resummed using the renormalization group.
The coefficients have already been computed
to one-loop order [\ref{HiedaSuzuki}], but here it suffices to
know that
\equation{
  c_1(t)=1+\rmO(g^2),
  \qquad
  c_2(t)=\rmO(g^4),
  \enum
}
and that the anomalous-dimension matrix of the fields (17),(18) is of
order $g^4$. Together with the renormalization group, these properties
imply
\equation{
  \lim_{t\to0}q(t,x)=\qRGI(x),
  \enum
}
where $\qRGI(x)$ is the renormalization-group-invariant charge density
(the normalization conventions are as in ref.~[\ref{DRqtop}]).

\section 6

{\bf 6.}~At this point, the equality of the gradient-flow and the density-chain
expressions for the topological susceptibility can be shown in a few lines.
First note that
\equation{
  \langle Q_t^2\rangle_{\rm c}=
  \langle Q_tQ\rangle_{\rm c}
  \enum
}
in view of the independence of the charge $Q_t$ on the flow
time $t$, the small flow-time limit (21) and the fact that
$\qRGI(x)$ coincides with $q(x)$ up to a term
proportional to the divergence of the singlet axial current.
The correlation function $\langle Q_tQ\rangle_{\rm c}$
does not require renormalization and may be represented through
\equation{
  \langle Q_tQ\rangle_{\rm c}={1\over3i}\delta^A_{\lambda}
  \langle Q_t\rangle,
  \enum
}
where $\lambda$ is given by eq.~(11).

For the same reasons ($t$-independence, etc.),
\equation{
  \langle Q_t\rangle=\langle Q\rangle=-{1\over2i}
  \delta^A_{\eta}\FR(M),
  \enum
}
and the combination of eqs.~(22)-(24) then implies
\equation{
  \langle Q_t^2\rangle_{\rm c}=
  {1\over6}\delta^A_{\lambda}\delta^A_{\eta}\FR(M).
  \enum
}
After setting $M$ to $\Mbar$ and recalling eq.~(15),
$\chifl$ is
thus seen to coincide with $\chidc$.

\section 7

{\bf 7.}~If there are less than $5$ flavours of sea quarks,
valence quarks are included in the theory and
the equivalence of the density-chain and
the gradient-flow representations
of the topological susceptibility can then again be proved
following the lines of the previous sections.
More precisely, the theory is set up with
$\Nf$ fermionic and $\Nb\leq\Nf$ bosonic quark fields
with mass matrices $M$ and $\Mt$, respectively,
such that the effects of the first $\Nb$ fermionic quark fields
are canceled if the mass matrices match. The
net number of sea quarks is then $\Nf-\Nb$, while $\Nb$
can be arbitrarily large and is assumed to be at least $5$
in the following.

In this theory,
there are separate flavour symmetries acting on the two types of
quark fields and additionally a supersymmetry that mixes
the bosonic with the first $\Nb$ fermionic fields.
These symmetries imply that the counterterm required to
cancel the singularities of the free
energy $\F(M,\Mt)$ must be of the form
\equation{
  \Delta\F(M,\Mt)=V\biggl[{k_0\over a^4}+
  {k_1\over a^2}\bigl(\tr\{M^{\dagger}M\}-\tr\{\Mt^{\dagger}\Mt\}\bigr)
  \noenum
  \next{2.0ex}
  \quad
  +k_2\bigl(\tr\{M^{\dagger}M\}-\tr\{\Mt^{\dagger}\Mt\}\bigr)^2
  +k_3\bigl(\tr\{(M^{\dagger}M)^2\}-\tr\{(\Mt^{\dagger}\Mt)^2\}\bigr)
  \biggr].
  \enum
}
Moreover, although there is now a second flavour-singlet axial current,
$\tilde{A}^s_{\mu}(x)$, constructed from the
bosonic quark fields, the symmetries imply that
there are still only two fields of dimension $4$,
\equation{
  \phiR{1}(x)=q(x)+\varzqa
  \drv{\mu}\bigl(A^s_{\mu}(x)+\tilde{A}^s_{\mu}(x)\bigr),
  \enum
  \next{2.0ex}
  \phiR{2}(x)=\varza\drv{\mu}\bigl(A^s_{\mu}(x)+\tilde{A}^s_{\mu}(x)\bigr),
  \enum
}
which can occur in the small flow-time expansion (16).

The density-chain expression for
the topological susceptibility can now be shown to coincide with
the gradient-flow expression
by repeating the steps taken in the
case of the theory with $5$ or more flavours of sea quarks.
In this derivation, the bosonic quark fields play a spectator r\^ole and
the associated mass matrix $\Mt$ can be set to the
diagonal matrix with diagonal elements $m_1,\ldots,m_{\Nb}$
from the beginning.

\section 8

{\bf 8.}~Unlike the chiral anomaly, which is a robust property of the
local field algebra, topological concepts are not obviously meaningful
in quantum field theory, since the fluctuations of the fields integrated over
in the functional integral are typically highly irregular.
Along the Yang--Mills gradient flow, the high-frequency fluctuations of the
gauge field in QCD are damped consistently with the renormalization group,
which permits the division of field space in sectors of
fixed topological charge to be given a well-defined and
regularization-independent meaning [\ref{WilsonFlow}].

The fact that the associated topological susceptibility coincides
with the density-chain expression (2) shows that these sectors
are related to the chiral anomaly as suggested by semi-classical studies,
where the background fields are assumed to be smooth.
While this has long been suspected to be the case,
proving the equality of the two expressions for the susceptibility
requires control over the theory at the non-perturbative level,
which is, at present, only achievable through
working hypotheses such as the assumption that
the continuum limit of lattice QCD exists and that
the small flow-time expansion (16) holds beyond perturbation theory.

\vskip1ex
I wish to thank Leonardo Giusti and Peter Weisz for
correspondence on the subject as well as for
their critical comments on a first version of this paper
and on various preparatory notes.

\beginbibliography


\bibitem{WilsonFlow}
M. L\"uscher,
{\it Properties and uses of the Wilson flow in lattice QCD},
JHEP 1008 (2010) 071 [Erratum: {\it ibid.} 1403 (2014) 092]

\bibitem{RenFlow}
M. L\"uscher, P. Weisz,
{\it Perturbative analysis of the gradient flow in non-Abelian gauge
theories},
JHEP 1102 (2011) 051

\bibitem{SuzukiRen}
K. Hieda, H. Makino, H. Suzuki,
{\it Proof of the renormalizability of the gradient flow},
Nucl. Phys. B918 (2017) 23


\bibitem{ChiWII}
L. Giusti, G. C. Rossi, M. Testa,
{\it Topological susceptibility in full QCD with Ginsparg--Wilson fermions},
Phys. Lett. B587 (2004) 157

\bibitem{ChiDC}
M. L\"uscher,
{\it Topological effects in QCD and the
problem of short distance singularities},
Phys. Lett. B593 (2004) 296


\bibitem{CeEtAl}
M. C\`e, C. Consonni, G. P. Engel, L. Giusti,
{\it Non-Gaussianities in the topological charge distribution of
the SU(3) Yang--Mills theory},
Phys. Rev. D92 (2015) 074502


\bibitem{ETMTop}
C. Alexandrou et al.,
{\it Topological susceptibility from twisted mass fermions
using spectral projectors and the gradient flow},
Phys. Rev. D97 (2018) 074503


\bibitem{GW}
P. H. Ginsparg, K. G. Wilson,
{\it A remnant of chiral symmetry on the lattice},
Phys. Rev. D25 (1982) 2649

\bibitem{DWF}
D. B. Kaplan,
{\it A method for simulating chiral fermions on the lattice},
Phys. Lett. B288 (1992) 342

\bibitem{GWindex}
P. Hasenfratz, V. Laliena, F. Niedermayer,
{\it The index theorem in QCD with a finite cutoff},
Phys. Lett. B427 (1998) 125

\bibitem{PerfectDiracOperator}
P. Hasenfratz,
{\it Prospects for perfect actions},
Nucl. Phys. B (Proc. Suppl.) 63 (1998) 53;
{\it Lattice QCD without tuning, mixing and current renormalization},
Nucl. Phys. B525 (1998) 401

\bibitem{NeubergerOperator}
H. Neuberger,
{\it Exactly massless quarks on the lattice},
Phys. Lett. B417 (1998) 141;
{\it More about exactly massless quarks on the lattice},
Phys. Lett. B427 (1998) 353

\bibitem{ExactChSym}
M. L\"uscher,
{\it Exact chiral symmetry on the lattice and the Ginsparg--Wilson relation},
Phys. Lett. B428 (1998) 342

\bibitem{GWreview}
F. Niedermayer,
{\it Exact chiral symmetry, topological charge and related topics},
Nucl. Phys. Proc. Suppl. 73 (1999) 105


\bibitem{Mainz}
M. L\"uscher,
{\it Future applications of the Yang--Mills gradient flow in lattice QCD},
PoS LATTICE2013 (2014) 016

\bibitem{HiedaSuzuki}
K. Hieda, H. Suzuki,
{\it Small flow-time representation of fermion bilinear operators},
Mod. Phys. Lett. A 31 (2016) 38


\bibitem{GiustiEtAlI}
L. Giusti, G. C. Rossi, M. Testa, G. Veneziano,
{\it The $U_A(1)$ problem on the lattice with Ginsparg--Wilson fermions},
Nucl. Phys. B628 (2002) 234

\bibitem{DRqtop}
M. L\"uscher, P. Weisz,
{\it Renormalization of the topological charge density in QCD
with dimensional regularization},
Eur. Phys. J. C81 (2021) 519

\endbibliography

\bye